\documentclass[preprint2,times,tighten]{aastex6}
\usepackage{amsmath,amstext}
\usepackage[figure,figure*]{hypcap}
\usepackage{newtxmath} 
\usepackage{natbib}

\shorttitle{NGC 6744: A Milky Way twin with a LINER}
\shortauthors{ da Silva, Steiner, \& Menezes}

\begin{document}

\title[NGC 6744 - A nearby Milky Way twin with a very low-luminosity AGN]{NGC 6744 - A nearby Milky Way twin with a very low-luminosity AGN}
\author{Patr\'icia da Silva$^1$, J. E. Steiner$^1$, R.B. Menezes$^{1,2}$}
\affiliation{$^1$Instituto de Astronomia, Geof\'isica e Ci\^encias Atmosf\'ericas, Departamento de Astronomia, Universidade de S\~ao Paulo,  05508-090, SP, Brazil\\
$^2$Centro de Ci\^encias Naturais e Humanas, Universidade Federal do ABC, 09210-580, SP, Brazil \\
\textit{Received 2017 December 2; revised 2018 May 10; accepted 2018 May 16; published 2018 July 6}}
\email{p.silva2201@gmail.com \\ joao.steiner@iag.usp.br \\ roberto.menezes@iag.usp.br}

\begin{abstract}

NGC 6744 is the nearest and brightest south-hemisphere galaxy with a morphological type similar to that of the Milky Way. Using data obtained with the Integral Field Unit of the Gemini South Multi-Object Spectrograph, we found that this galaxy has a nucleus with LINER (Low Ionization Nuclear Emission Line Region) surrounded by three line emitting regions. The analysis of the Hubble Space Telescope archival images revealed that the nucleus is associated with a blue compact source, probably corresponding to the active galactic nucleus (AGN). The circumnuclear emission seems to be part of the extended narrow line region of the AGN. One of these regions, located $\sim$1$\arcsec$ southeast of the nucleus, seems to be associated with the ionization cone of the AGN. The other two regions are located $\sim$1$\arcsec$ south and $\sim$0$\arcsec$\!\!.6 northeast of the nucleus and are not aligned with the gaseous rotating disk. Spectral synthesis shows evidence that this galaxy may have gone through a merger about one billion years ago. On the basis of the kinematic behavior, we found a gaseous rotating disk, not co-aligned with the stellar disk. Given the relative degree of ionization and luminosities of the nuclear and circumnuclear regions, we suggest that the AGN was more luminous in the past and that the current circumnuclear emissions are echoes of that phase.

\end{abstract}


\section{Introduction}

NGC 6744 is an SAB(r)bc galaxy that belongs to the Pavo group and is located at 8.5 Mpc (Nasa Extragalactic Database- NED \footnote{The NASA/IPAC Extragalactic Database (NED) is operated by the Jet Propulsion Laboratory, California Institute of Technology, under contract with the National Aeronautics and Space Administration.}). It has a gas ring, with 3$\arcmin$\!\!.3 ($\sim$ 8.2 kpc) of diameter, where two spiral arms begin, as observed in H\textsc{ i} images \citep{ryder}. 

\begin{figure*}
\begin{center}
 \includegraphics[scale=0.6]{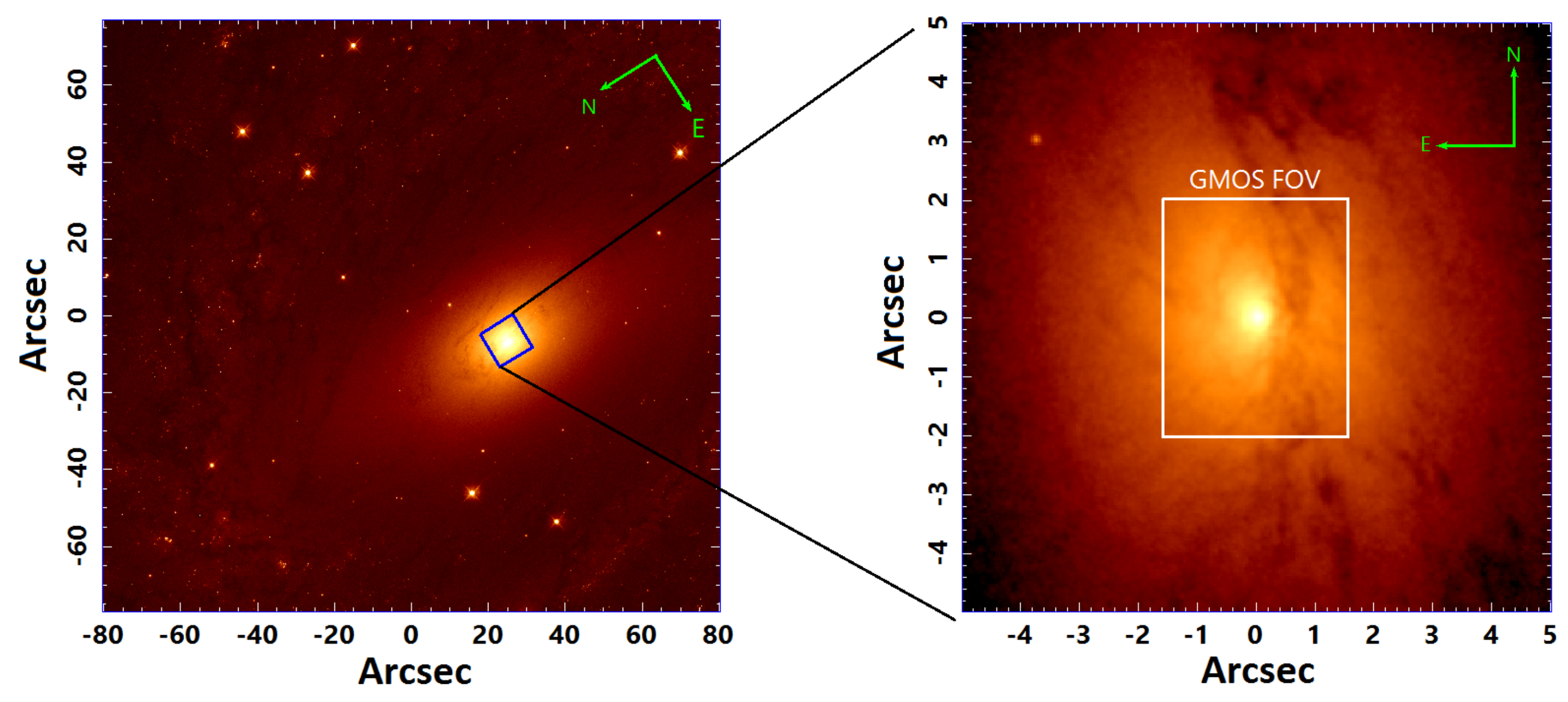}
  \caption{\textit{HST}/WFC3 images, in the F555W filter, of NGC 6744. In order to have an idea of the FOV of the GMOS observations, the figure on the left shows an image of all \textit{HST} FOV, with a little blue square on the central region, whose magnification is shown on the right with the GMOS FOV limits and its orientation. The size of the rectangle of GMOS FOV on x-axis is $\sim$ 130 pc and on y-axis is $\sim$ 167 pc. \label{hst_tot}}
\end{center}
\end{figure*}

Its bulge is classic, very similar to an elliptical galaxy, with a S\'ersic index of 3.1, measured on images obtained with the Spitzer telescope in 3.6 $\mu$m \citep{fisher}. In this respect, this galaxy is different from the Milky Way, which has a peanut-shaped pseudo-bulge. It has a tenuous trace of dust toward the center. The galaxy has a star forming-ratio of 6.8 M$_\sun$/year \citep{ryder2}.

According to \citet{vacelli}, NGC 6744 has a nucleus whose emission line ratios are typical of low ionization nuclear emission line regions (LINERs), which are often active galactic nuclei (AGNs) that present an emission spectrum with a lower ionization degree when compared to Seyferts \citep{heckman80}. \citet{ferland} and \citet{halpern} demonstrated that photoionization by an AGN, with a low ionization parameter, results in the observed lower ionization spectrum of LINERs; though, other scenarios may be applicable to some objects (see \citealt{ho08} for a review).

Infrared studies (LW2 and LW3 filters of Spitzer telescope) have shown that there is little emission in the region between the nucleus and the ring and this area is dominated by old stellar populations \citep{roussel}. Besides that, in this region, there is not much star formation, which may be due to the orbital resonance that ejected the gas \citep{ryder3}. 

Our group is conducting the Deep IFS View of Nuclei of Galaxies (DIVING3D) survey ( J. E. Steiner et al. 2018, in preparation), which has the goal of observing the nuclear region of all southern galaxies brighter than B=12.0. The observations are being taken, in the optical, with the integral field unit (IFU) of the Gemini Multi-Object Spectrograph (GMOS), on the Gemini-South telescope. One of the goals is to analyze in detail the subsample with a morphological type similar to that of the Milky Way (SABbc or SBbc - which we call Milky Way twins). NGC 6744 is the brightest (B = 9.24 - NED) object of this subsample and, consequently, deserves special attention. 

In this work, we analyze the data cube of the central region of NGC 6744, in order to characterize the nature of the emission from this region and to study the kinematics of the gas and stars. In Section \ref{sec2}, we show details of the observations, of the reduction processes, and of the data analysis. Section \ref{sec3} presents the analysis of the emitting spectrum of different spatial regions of the data cube. In sections \ref{sec4} and \ref{sec6}, we present the analysis of the kinematics of the gas and stars, respectively. We show the results of the spectral synthesis applied to the spectra of two distinct regions in Section \ref{sec5}. In Section \ref{hst}, we present the analysis of the \textit{Hubble Space Telescope} (\textit{HST}) images of the central region of NGC 6744. We discuss all the obtained results in Section \ref{sec8}. Lastly, Section \ref{sec9} summarizes the main conclusions of this work.

\section{Observations and data reduction} \label{sec2}

The observations were taken on 2014 May 8, with GMOS/IFU, installed on the Gemini-South telescope, in one slit mode (GS-2014A-Q-5). Three 815 s exposures were made, with position angles (PA) equal to 0$\degr$ and with fields of view (FOVs) of 5$\arcsec \times $3$\arcsec$\!\!.5. The grating used for this observation was R831+G5322, with the central wavelength of 5850\AA, and the resulting spectral resolution was R=4340. The seeing, determined from the acquisition image (observed at $\lambda$ = 6300 \AA), was FWHM $\sim$ 0$\arcsec$\!\!.55. Fig.~\ref{hst_tot} shows the GMOS FOV on an \textit{HST} image, obtained with the WFC3 camera. One can notice that the observation area is restricted only to the $\sim$ 130 $\times$ 167 pc of the central region of the bulge of NGC 6744.

The reduction was performed in the \textsc{iraf} environment, and it started with the determination of the trim and bias subtraction. After that, the cosmic-ray removal was applied using the \textsc{lacos} routine \citep{van01}. Then spectra were extracted and corrected for gain variations between the pixels (with response curves obtained with GCAL-flat images). Corrections for gain variations between fibers and asymmetric illumination patterns of the instrument were also performed (using twilight-flat images). The spectra were wavelength calibrated (using CuAr lamp images) and the sky subtraction was applied (using the mean spectrum of the sky FOV, located at 1$\arcmin$ from the science FOV). Corrections of the atmospheric extinction and telluric absorption removal were then applied. After the flux calibration, three data cubes were created. These cubes have spatial pixels (spaxels) with  0$\arcsec$\!\!.05 $\times$ 0$\arcsec$\!\!.05 and their spectral coverage is from 4804\AA\ to 6865\AA. 

The data cubes' treatment involved the use of some scripts, which were written by our group, in IDL (Interactive Data Language). This treatment consisted of the correction of the differential atmospheric refraction, the combination of the data cubes to determine a median (in order to obtain only one data cube), Butterworth spatial filtering \citep{gwoods}, and instrumental fingerprint removal (see \citealt{rob1,rob2}). Lastly, we applied the Richardson-Lucy deconvolution (\citealt{rich} and \citealt{lucy}), with 10 iterations, where the wavelength variation of the FWHM of the point spread function (PSF) was estimated from the data cube of the standard star, and the FWHM value of a point-like source from the acquisition image was used to determine the PSF in a specific wavelength. After comparing with an \textit{HST} image in the F555W (V) filter convolved with Gaussian PSFs, we estimate that the FWHM of the PSF of the treated data cube is 0$\arcsec$\!\!.43, which corresponds to 17.7 pc.

\begin{figure*}
\begin{center}
 \includegraphics[scale=0.6]{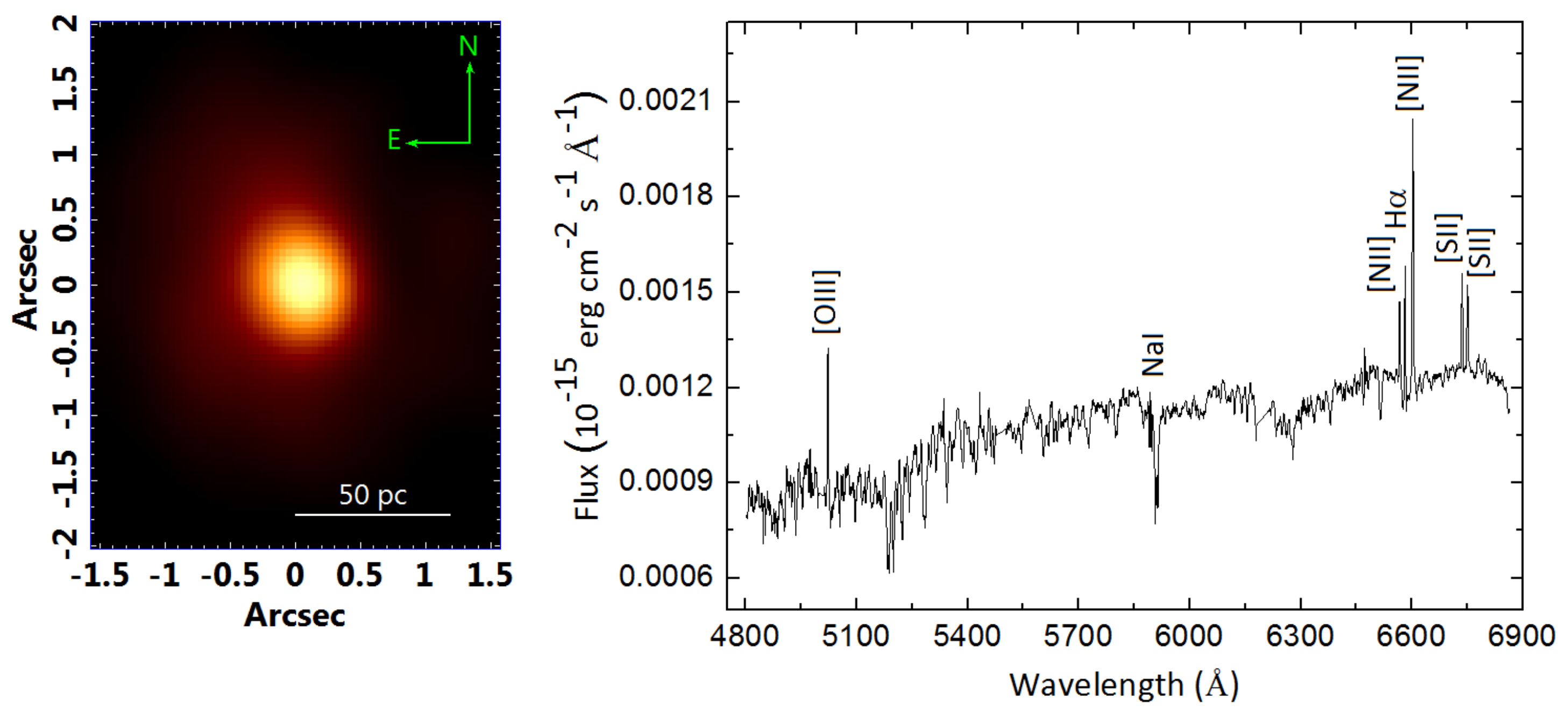}
  \caption{Image of the data cube, collapsed along the spectral axis, observed with GMOS IFU, together with its mean spectrum. The NE orientation and the 50 pc scale are also shown in the image. \label{cubecolapsed}}
\end{center}
\end{figure*}

Fig.~\ref{cubecolapsed} shows the resulting image of the sum of all images of the data cube after the treatment and also its mean spectrum. This mean spectrum reveals the following emission lines: [O \textsc{iii}]$\lambda$5007, [O \textsc{i}]$\lambda$6300, [N \textsc{ii}]$\lambda\lambda$6548, 6584, H$\alpha$ e [S \textsc{ii}]$\lambda\lambda$6716, 6731. Those lines are emitted, mostly, in the central region and are quite narrow.

\section{Nuclear and circumnuclear emission line properties} \label{sec3}

\subsection{Spectral synthesis and subtraction of stellar continuum}

In order to study the emission lines, both in terms of the regions where they are emitted and in terms of the emission line ratios and profiles, it is necessary to analyze the data cube without stellar continuum. For that, we performed a spectral synthesis (\textsc{starlight} - \citealt{starlight}) of the data cube with masked emission lines using a base with stellar population spectra based on the Medium-resolution Isaac Newton Telescope Library of Empirical Spectra - MILES \citep{blazquez}. From our previous experience, a minimum signal-to-noise ratio (S/N) of 10 is necessary to obtain reliable results from the spectral synthesis. We verified that the S/N of the continuum, between 5804\AA\ and 5835\AA,  of the data cube is 11$\leqslant$ S/N $\leqslant$ 54, with the values increasing toward the center; therefore, all the parameters provided by the spectral synthesis seem to be reliable.

We built a data cube with the synthetic stellar spectra corresponding to the fits provided by the spectral synthesis. Such a data cube was subtracted from the original one, which resulted in a data cube with emission coming mainly from gas, which we refer to here as the gas data cube. 

\subsection{The line emitting regions}

\begin{figure*}
\begin{center}
 \includegraphics[scale=0.55]{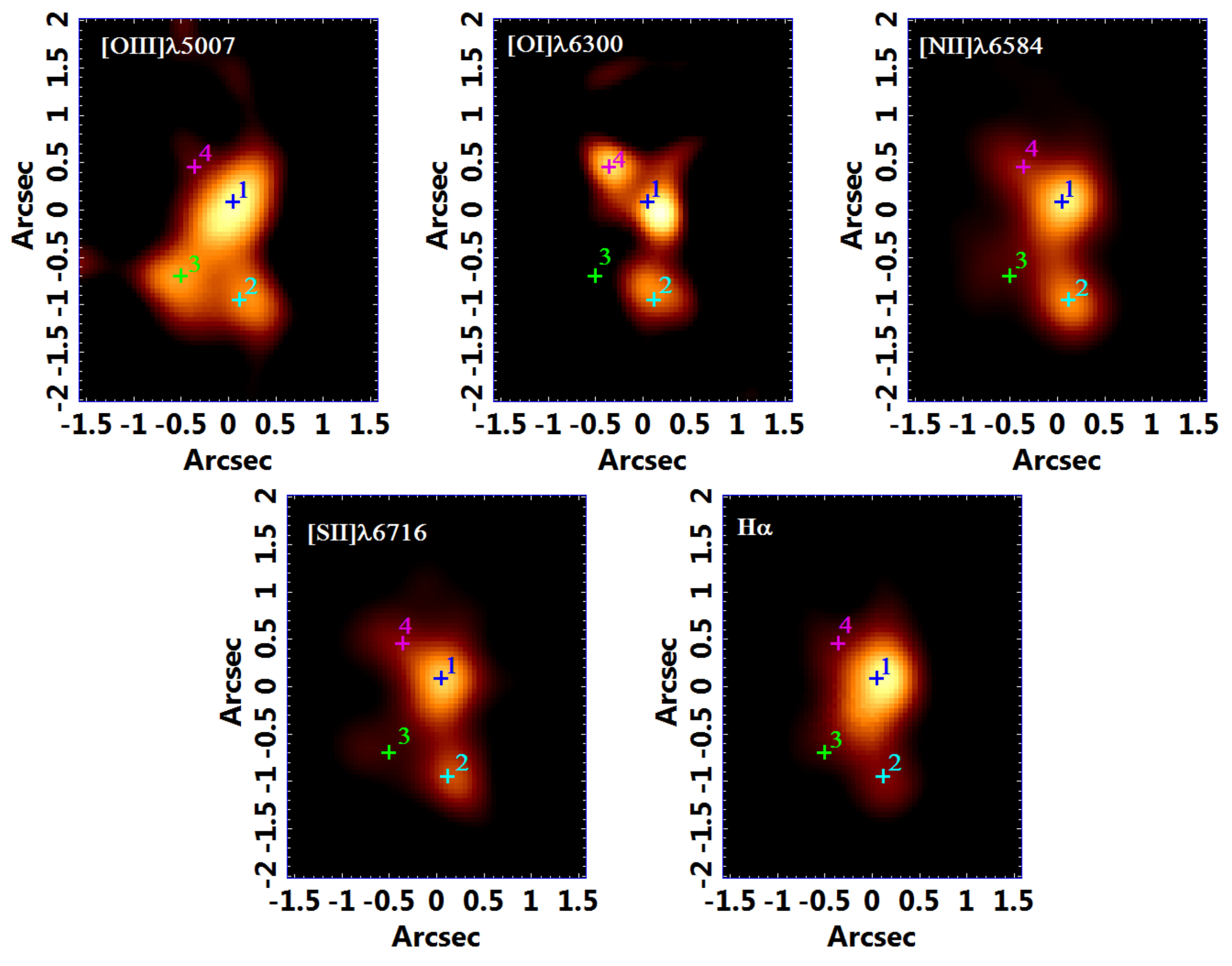}
  \caption{Images of the main emission lines of the NGC 6744 data cube after the stellar continuum subtraction. All the images have indications of the centroids of all emitting regions. The positions of Regions 1 and 2 were determined from the [N \textsc{ii}]$\lambda$6584 image, the position of Region 3 from [O \textsc{iii}]$\lambda$5007, and the position of  Region 4 from [O \textsc{i}]$\lambda$6300 image. Note that Region 3 does not appear in evidence in the images (except in the [O \textsc{iii}]$\lambda$5007 image) and, in the H$\alpha$ image, only Region 1 appears in evidence. Region 4, which is very bright in the [O \textsc{i}]$\lambda$6300 image (together with Regions 1 and 2), seems to be very diluted in the other images. Besides that, there is an extension of Region 1 toward Region 3 in the [O \textsc{iii}]$\lambda$5007 image.\label{6images}}
\end{center}
\end{figure*}

\begin{figure*}
\begin{center}
   \includegraphics[scale=0.75]{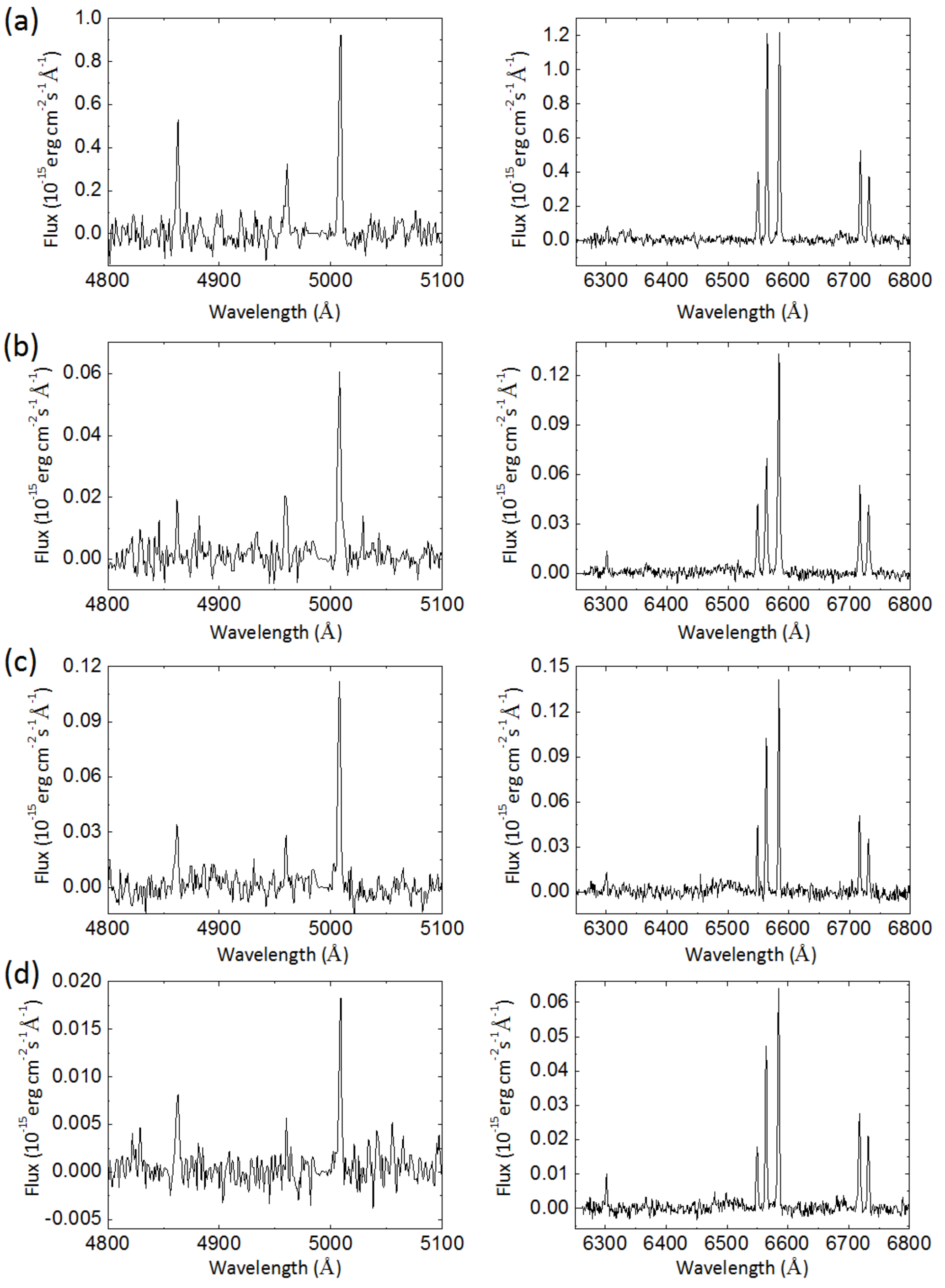}
  \caption{Blue and red parts of the emission line spectra of (a) Region 1, (b) Region 2, (c) Region 3, and (d) Region 4, showing the main emission lines. \label{espectros4}}
\end{center}
\end{figure*}

From the gas data cube, it was possible to obtain images of some emission lines (see Fig.~\ref{6images}). The main source of all emission lines is called Region 1. The position of this region is very close to the center of the FOV, which coincides approximately with the center of the bulge (the galactic nucleus). A second region (Region 2) appears in all emission line images and is located $\sim$ 0\arcsec\!\!.9 south from the nucleus. Region 3, more visible in [O \textsc{iii}]$\lambda$5007 image, is $\sim$ 1\arcsec to the southeast from Region 1. Lastly, Region 4, more easily detected in [O \textsc{i}]$\lambda$6300 image, is located $\sim$ 0\arcsec\!\!.6 northeast from the nucleus. The nucleus appears elongated in the [O \textsc{iii}]$\lambda$5007 image, with PA $\sim$ - 53\degr (toward Region 3). Table \ref{tabeladist} shows the coordinates of the centroids of each region, assuming that the centroid of Region 1 is at $X_C =0 \arcsec$ and $Y_C =0 \arcsec$. The projected distances ((\textit{D}) in parsecs) of each object relative to Region 1 are also presented.

\begin{table}
\centering
\caption{Projected Distances (\textit{D}) in Parsecs of Regions 2, 3, and 4 Relative to Region 1 ($X_C =0 \arcsec$ and $Y_C =0 \arcsec$).  $X_C$ and $Y_C$ are the central coordinates of each region. The uncertainty for \textit{D} was calculated taken into account only the instrumental uncertainties.}
\label{tabeladist}
\begin{tabular}{cccc}
\hline
Regions & $X_C$($\arcsec$) & $Y_C$($\arcsec$) & $D$ (pc)         \\ \hline
2       & -0.1             & 0.9                & 41.2 $\pm$ 1.5 \\
3       & 0.5              & 0.8              & 40.5 $\pm$ 1.5 \\ 
4       & 0.4              & -0.5              & 22.2 $\pm$ 1.5 \\ \hline
\end{tabular}
\end{table}

\begin{table*}
\centering
\caption{ FWHM of the [N \textsc{ii}]$\lambda$6584 Line (Corrected for the Instrumental Resolution), H$\alpha$ Luminosity, Eletronic Density Intervals (Obtained from [S \textsc{ii}]$\lambda$6716/[S \textsc{ii}]6731 Ratio), and Upper Limit for the Temperature (Obtained from the Ratio of [N \textsc{ii}]$\lambda$5755 and [N \textsc{ii}]$\lambda$6548 + [N \textsc{ii}]$\lambda$6584 Lines) of the Observed Regions.}
\label{tableetc}
\begin{tabular}{ccccc}
\hline
Region                                                   & 1                      & 2                      & 3               & 4             \\ \hline
FWHM ({[}N \textsc{ii}{]}$\lambda$6584) in km s$^{-1}$ & 152 $\pm$ 10                & 180 $\pm$ 13                    & 123 $\pm$ 15             & 150 $\pm$ 17           \\
Luminosity of H$\alpha$ ($10^{36}$erg s$^{-1}$)          & 33.9 $\pm$ 0.8         & 2.62 $\pm$ 0.15        & 2.97 $\pm$ 0.18 & 1.4 $\pm$ 0.4 \\
Electronic density ( $n_e$ in cm$^{-3}$)                     & 258 $\geq$ d $\geq$ 81 & 409$\geq$ d $\geq$ 230 & d $\leq$ 64     & d $\leq$ 170  \\
Upper limit for temperature (K)                        & 40305                  & 8048                   & 9367            & 6539          \\ \hline
\end{tabular}
\end{table*}

\subsection{Diagnostic diagrams}

 \begin{figure*}
\begin{center}
     \includegraphics[scale=0.65]{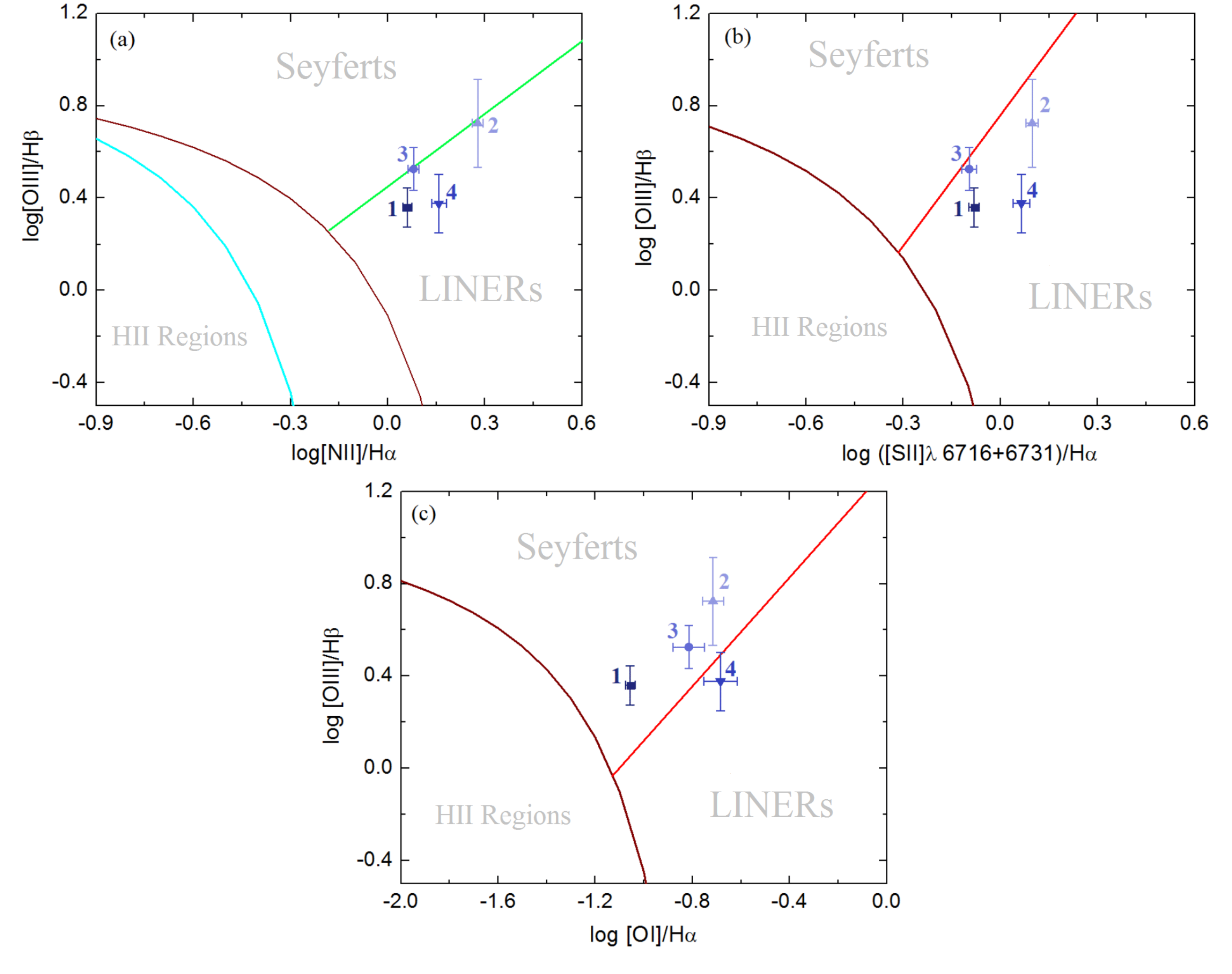}
  \caption{Diagnostic diagrams with the emission line ratios of the regions observed in the data cube of the central region of NGC 6744, using the values of Table \ref{tablerazao}. The four regions are represented by points in violet color variations. Each region is identified in the graph with its respective number next to each point. The wine fit in the (a), (b), and (c) diagrams shows the upper limit for the ionization by a starburst obtained by \citet{kewley1}. The cyan fit represents the division between H \textsc{ii} regions and AGNs obtained by \citet{kauffmann} and the red line, the division between LINERs and Seyferts in diagrams (b) and (c) determined by \citet{kewley2}. The green line in diagram (a) is the division between LINERs and Seyferts determined by \citet{schawinski}. \label{diagramdiagn}}
  
\end{center}
\end{figure*}

\begin{table*}
\centering
\caption{Emission Line Ratios for the Four Observed Regions.}
\label{tablerazao}
\begin{tabular}{ccccc}
\hline
Emission Line Ratios                       & Region 1   & Region 2 & Region 3 & Region 4 \\ \hline
{[}O \textsc{iii}{]}$\lambda$5007/H$\beta$              & 2.3 $\pm$ 0.4     & 5.3 $\pm$ 1.9   & 3.4 $\pm$ 0.6   & 2.4 $\pm$ 0.6          \\
{[}N \textsc{ii}{]}$\lambda$6584/H$\alpha$              & 1.15 $\pm$ 0.03   & 1.90 $\pm$ 0.07 & 1.20 $\pm$ 0.05 & 1.44 $\pm$ 0.07          \\
({[}S \textsc{ii}{]}$\lambda$6716+6731)/H$\alpha$       & 0.83 $\pm$ 0.03   & 1.25 $\pm$ 0.05 & 0.80$\pm$ 0.04 & 1.16 $\pm$ 0.07           \\
{[}O \textsc{i}{]}$\lambda$6300/H$\alpha$               & 0.088 $\pm$ 0.004 & 0.192 $\pm$ 0.018 & 0.153 $\pm$ 0.021 & 0.21 $\pm$ 0.03          \\
{[}S \textsc{ii}{]}$\lambda$6716/{[}S \textsc{ii}{]}$\lambda$6731 & 1.26 $\pm$ 0.08   & 1.14 $\pm$ 0.06 & 1.49 $\pm$ 0.13 & 1.36 $\pm$ 0.11 \\ \hline       
\end{tabular}
\end{table*}

In order to characterize the nature of the emission of the four observed regions, we extracted spectra (from the gas data cube) of circular regions, whose central coordinates are in Table \ref{tabeladist} and whose extraction radius was equal to half of the FWHM of the PSF value: $0\arcsec\!\!.25$. Fig.~\ref{espectros4} shows the extracted spectra (the blue and red parts of the spectrum) of the four main regions.

 \begin{figure*}
\begin{center}
   \includegraphics[scale=0.78]{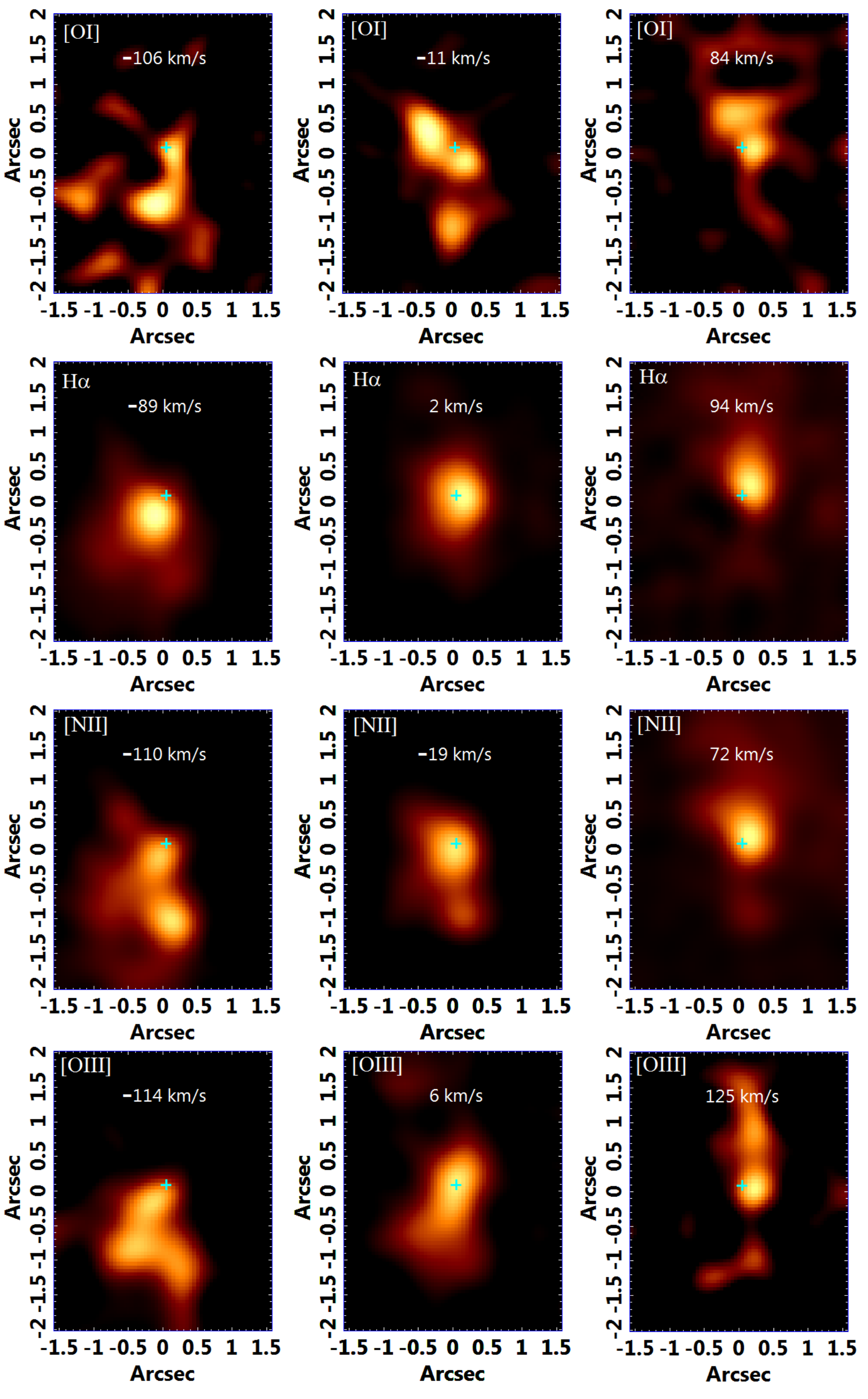}
  \caption{Channel maps of the following emission lines from the data cube of NGC 6744 after the stellar continuum subtraction: [O \textsc{i}]$\lambda$6300, H$\alpha$, [N \textsc{ii}]$\lambda$6584, and [O \textsc{iii}]$\lambda$5007. The interval of each image is equivalent to 1\AA. Region 1's central position is indicated by the cyan cross, and its size is 3$\sigma$. \label{channelmaps}}
\end{center}
\end{figure*}

\begin{figure*}
\begin{center}
   \includegraphics[scale=0.55]{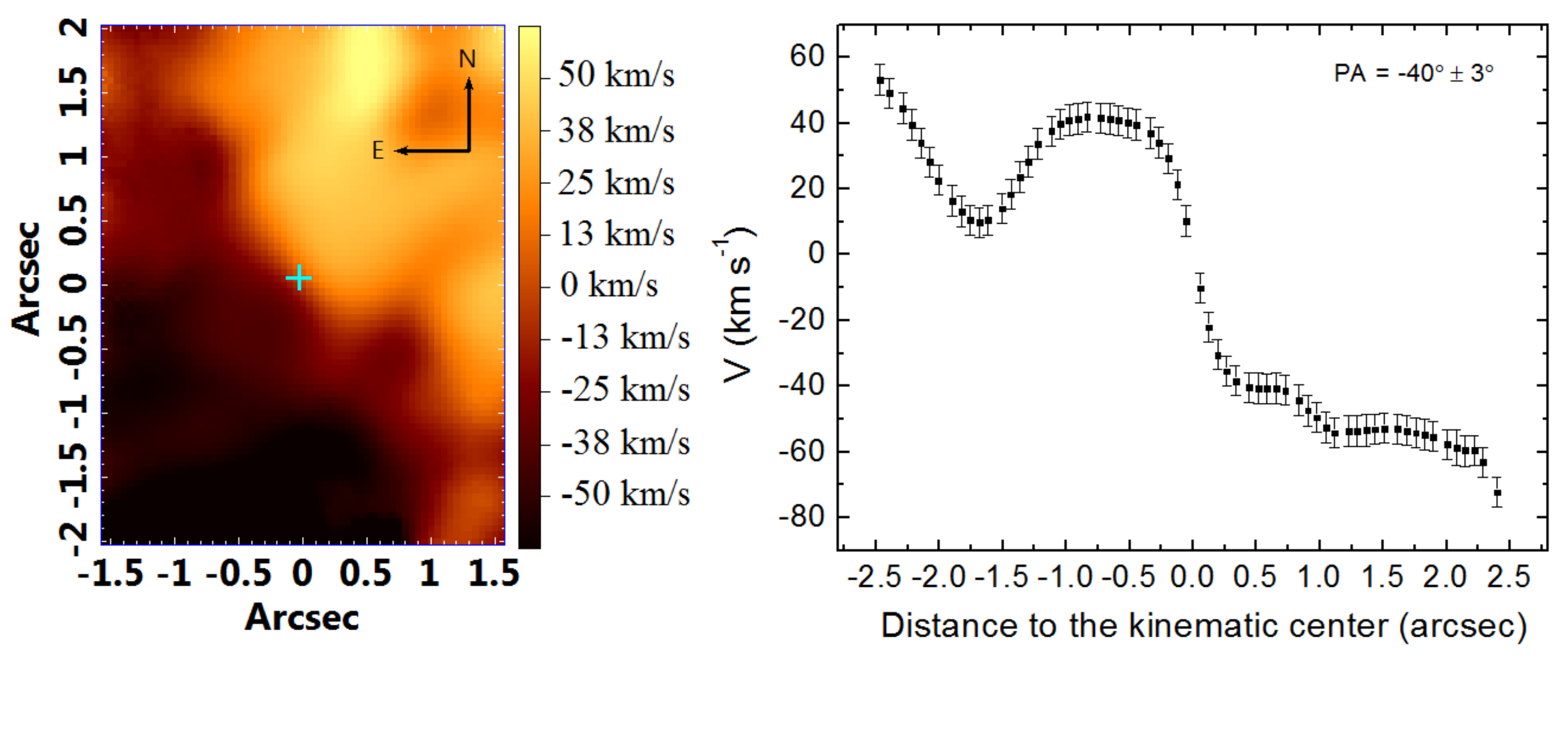}
  \caption{Velocity map of H$\alpha$ with its respective velocity curve, extracted along PA = -40$\degr$ $\pm$ 3$\degr$. The cyan cross represents the position of the kinematic center obtained for this map and its size indicates the uncertainty of 1$\sigma$. \label{vel_ha}}
\end{center}
\end{figure*}

\begin{figure*}
\begin{center}
   \includegraphics[scale=0.582]{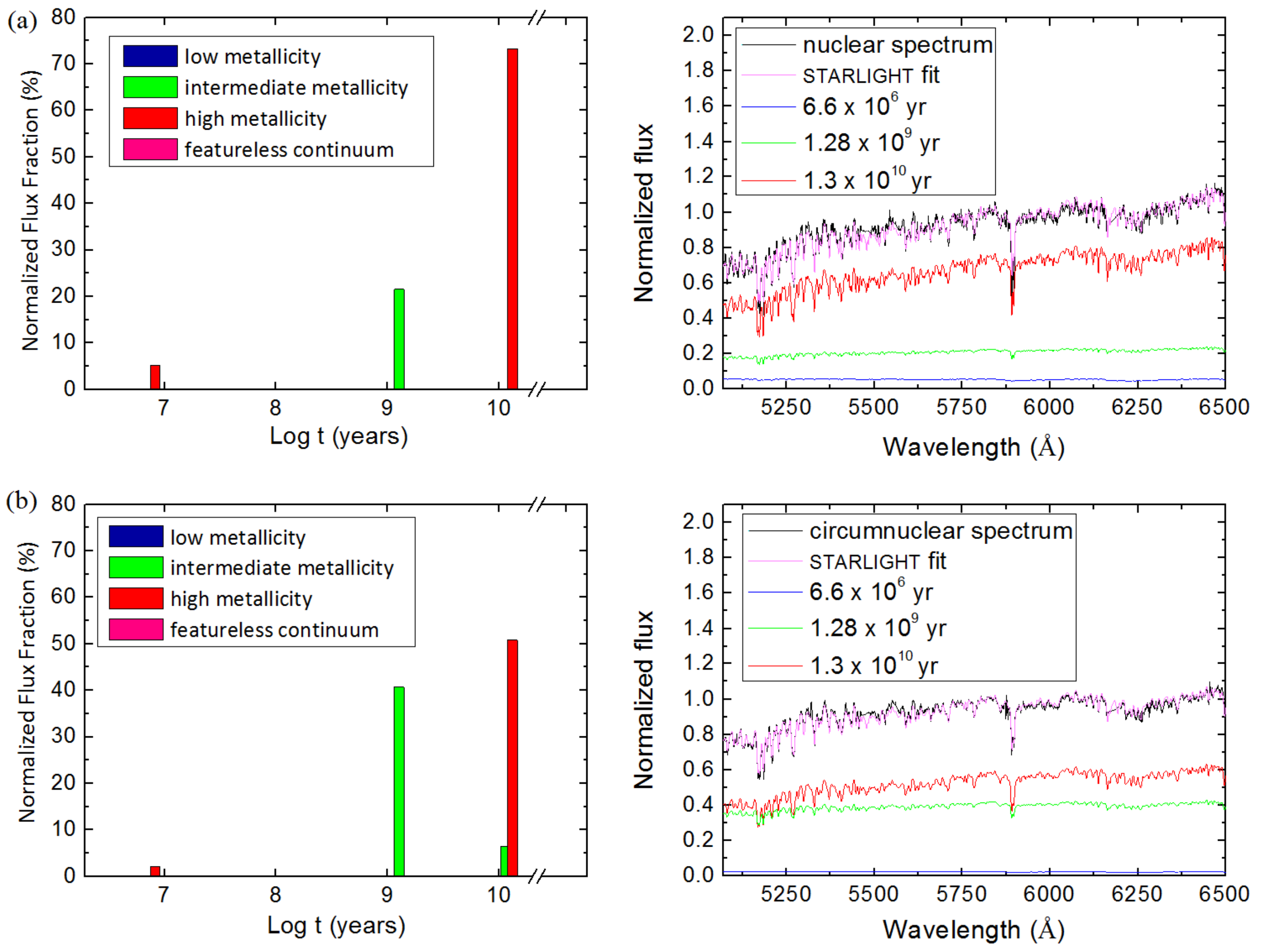}
  \caption{Spectral synthesis results of (a) the nuclear region spectrum and (b) the circumnuclear region spectrum. Both histograms show only stellar populations with high (0.02 e 0.05) and intermediate (4$\times$ $10^{-3}$ e 8$\times$ $10^{-3}$) metallicities, evidencing two significant stellar formation processes. Besides that, we can see that the higher flux fraction of the 10$^9$ year old stellar population is in the circumnuclear region. The extracted spectra, together with the corresponding \textsc{starlight} fits, in pink, and the spectra of the stellar populations required for the the fit are shown next to each histogram.\label{hist_star}}
\end{center}
\end{figure*}

The representative values of the FWHM corresponding to the emission lines of each region were calculated using  [N \textsc{ii}]$\lambda$6584 line, because this is the most intense line in all spectra. As we can see in Table \ref{tableetc}, the lines are quite narrow in all regions of the data cube. The emission line ratios were calculated using the integrated flux of each line of the extracted spectra (see Table \ref{tablerazao}).

The diagnostic diagrams (\citealt{baldwin,kewley1,kewley2,kauffmann}; Fig.~\ref{diagramdiagn}) were made based on the calculated emission line ratios and show that, in terms of ionization degree, all regions are compatible with each other, at the 3$\sigma$ level and all emission line ratios are compatible with those of Seyferts and LINERs. Regions 2 and 3 seem to have slightly higher ionization degrees when compared to the others. 

When analyzing the diagnostic diagrams together with the images of the emission lines presented in Fig.~\ref{6images}, we see that Region 3, which seems to have a higher ionization degree than Region 1, can be an ionization cone. This is because it appears in evidence only in the [O \textsc{iii}]$\lambda$5007 image, which is a higher ionization line, and also because Region 1 appears, in this same image, elongated toward Region 3. Region 4 appears in evidence only in the [O \textsc{i}]$\lambda$6300 image, suggesting that this might be a region of lower ionization.

\section{Gas Kinematics} \label{sec4}

From the gas data cube spectra, it is possible to study the gas kinematics using channel maps. In the present case, we made channel maps with 1\AA\ intervals (based on the fact that the emission lines were quite narrow) in each image, for each line, as is presented in Fig.~\ref{channelmaps}. For each emission line, we see that the first channel map represents the gas in blueshift, the second represents the gas with velocity close to zero, and the last represents the gas in redshift. The morphologies of the channel maps are similar and reveal a pattern consistent with a rotation around Region 1. However, some asymmetries in this pattern are also observed, like the fact that the PA value of the line of nodes in the redshifted gas region does not seem to be the same as that of the line of nodes in the blueshifted gas region (more clearly seen in the [O \textsc{iii}]$\lambda$5007 line channel maps). This may be explained by the fact that blueshifted gas with high velocities is associated with the possible ionization cone, thus being evidence of outflows of gas in that region.

We also note that Regions 2 and 3 are in blueshift and part of Region 4 has gas with low velocities in redshift.

From the H$\alpha$ velocity map (Fig.~\ref{vel_ha}) and its velocity curve, obtained from the Gaussian functions fitted to the H$\alpha$ emission line, we note that the probable gas rotation seems to be contained in the inner region and, in the FOV edges, there are some irregularities. The PA of the line of nodes is -40$\degr$ $\pm$ 3$\degr$. The kinematic center determined for this map is $X_C=0\arcsec\!\!.00 \pm 0\arcsec\!\!.11$, $Y_C=-0\arcsec\!\!.03 \pm 0\arcsec\!\!.11$, compatible, at the 1$\sigma$ level, with Region 1 centroid coordinates. Therefore, the inner gas is rotating around Region 1. We found that an amplitude/noise (A/N) ratio higher than 5 is necessary to obtain reliable radial velocity values from Gaussian functions fitted to the emission lines. Because the A/N ratio of the H$\alpha$ line is $\sim 5$ at the edges of the FOV and $\sim 60$ at the areas close to the stellar nucleus, we conclude that the values of the H$\alpha$ radial velocity are reliable, with a probable lower accuracy at the FOV edges, where the A/N ratio is close to the limit mentioned above.

The velocities of Regions 2, 3, and 4 relative to Region 1 are -54 $\pm$ 7 km s$^{-1}$, -55 $\pm$ 7 km s$^{-1}$, and -5 $\pm$ 5 km s$^{-1}$, respectively. The velocities of Regions 2 and 3 are similar (considering the uncertainties) and Region 4 seems to be at rest or in low velocity relative to Region 1.

We estimated values for the electronic density, based on the [S \textsc{ii}]$\lambda$6716/[S \textsc{ii}]$\lambda$6731 ratio and a gas temperature of 10$^4$ K. Upper limits for temperatures were calculated using the [N \textsc{ii}]$\lambda$5755, [N \textsc{ii}]$\lambda$6548 and [N \textsc{ii}]$\lambda$6584 line ratio (with an estimate for the upper limit of the integrated flux of [N \textsc{ii}]$\lambda$5755 line). Table \ref{tableetc} shows those values, as well as the values of the H$\alpha$ luminosity and the FWHM of the [N \textsc{ii}]$\lambda$6584 line of all observed regions. Note that, with the exception of Region 1, all of the regions have temperatures lower than $10^4$ K.

\section{Stellar archeology} \label{sec5}

In order to study the stellar populations in different areas of the FOV, we applied the spectral synthesis (using \textsc{starlight}) to two spectra of the data cube after treatment: one spectrum of the nuclear region (in this case, since Region 1 is the nuclear region, we used its spectrum) and one spectrum of the circumnuclear region (obtained by subtracting the spectrum of Region 1 from the total spectrum of the data cube). The spectral synthesis was applied as described in Section \ref{sec3}, but only to those two spectra and with an additional spectrum in the base, which is a power law with a spectral index of 1.5, representing a possible featureless continuum emission. 

The mean age uncertainty of the stellar populations (which is the weighted mean of the stellar population ages based on their flux fraction) was obtained with a Monte Carlo procedure and is of the order of 0.03 dex. The exact process for determining this uncertainty was the following: first, a synthetic stellar spectrum was obtained with the spectral synthesis applied to the mean spectrum of the data cube. A Gaussian distribution of the spectral noise was estimated from the result obtained by subtracting the synthetic stellar spectrum from the mean spectrum. Then, different Gaussian distributions were made, with the same width of the initial one; such distributions were added to the synthetic stellar spectrum mentioned above. The spectral synthesis was applied to each resulting spectrum and the uncertainty of the mean age was taken as the standard deviation of the mean ages obtained with those spectral synthesis applications.

\begin{figure*}
\begin{center}
   \includegraphics[scale=0.7]{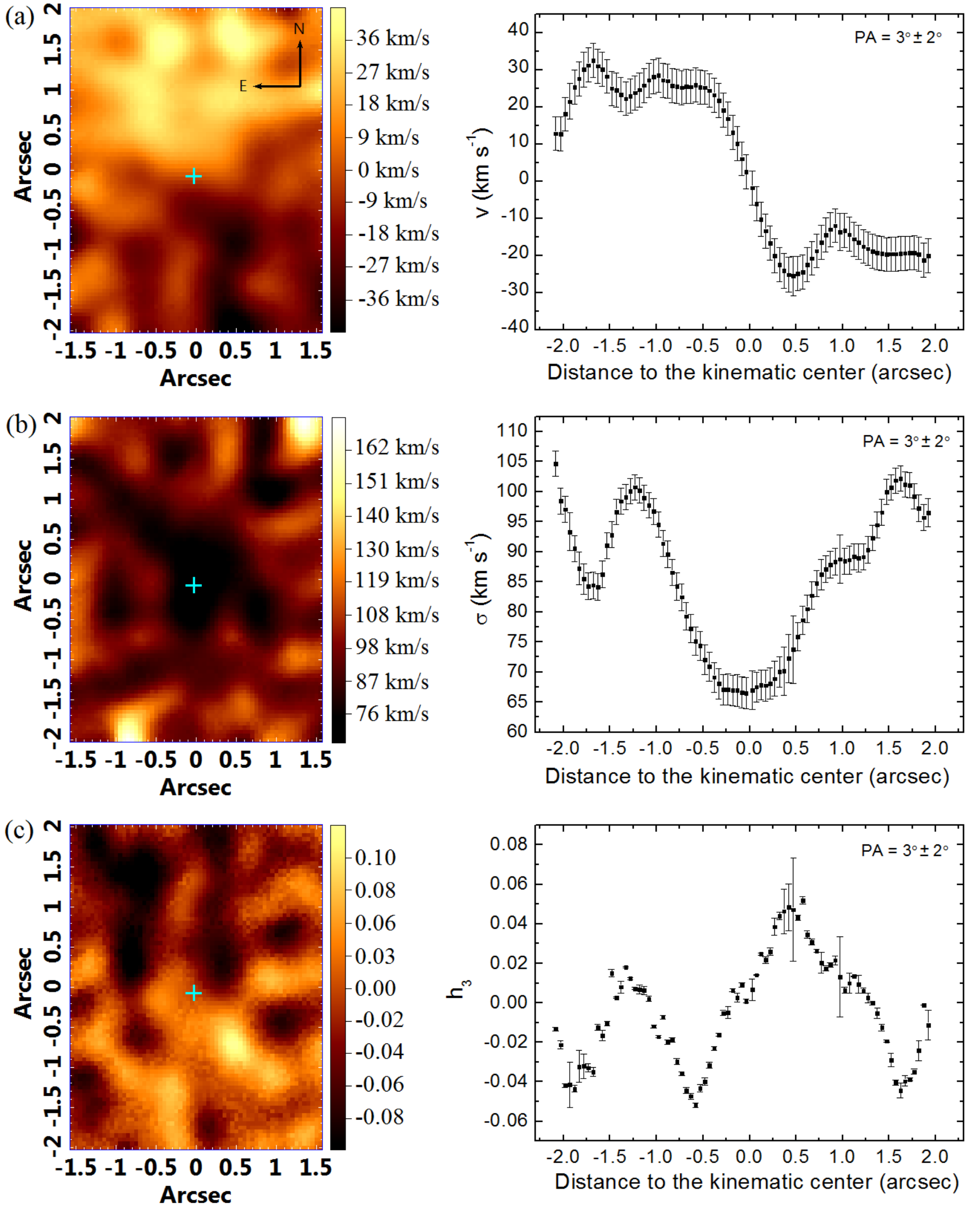}
  \caption{Maps of the following parameters obtained with pPXF: (a) $V_*$, (b) $\sigma_*$, and (c) $h_3$, with their respective curves extracted along the line of nodes, with PA indicated. The cyan cross represents the kinematic center position, obtained from the stellar velocity map and its size represents 1$\sigma$. The error bars in the curves of each map were determined from a Monte Carlo procedure, similar to the one described in Section \ref{sec5}. \label{ppxf}}
\end{center}
\end{figure*}

 \begin{figure*}
\begin{center}
   \includegraphics[scale=0.7]{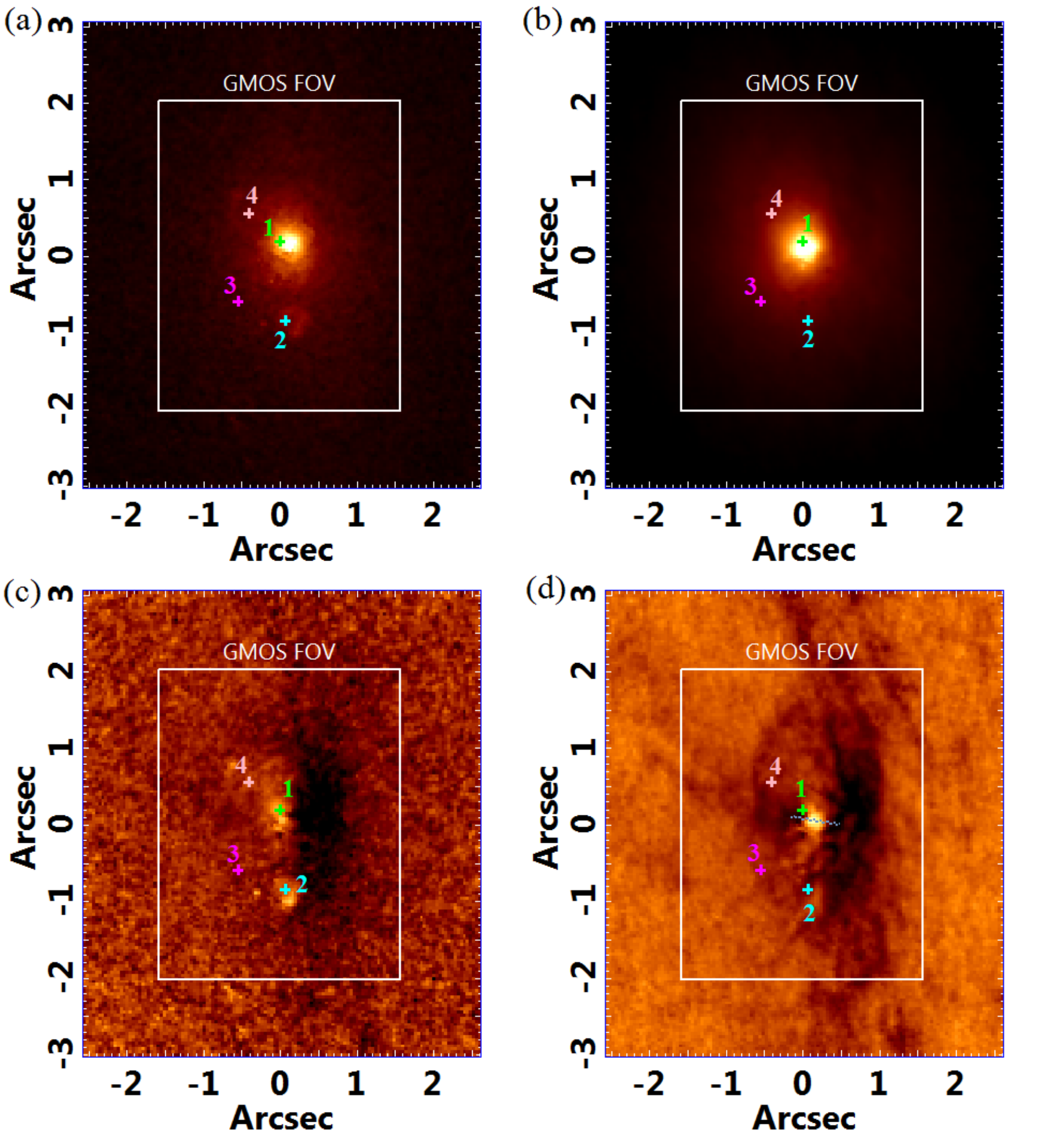}
  \caption{\textit{HST} images with the same orientation as the GMOS data cube analyzed in this work. The size of the GMOS data cube FOV is also indicated. All the images have the coordinates of the four emitting regions observed in the GMOS data cube indicated by crosses, and their sizes represent their uncertainties (3$\sigma$). The positions were determined assuming that the coordinates of Region 1 are compatible, at 3$\sigma$ level, with the center of the stellar bulge  (see Fig.~\ref{sourceregiao1}a). (a) H$\alpha$ + [N \textsc{ii}] emission image obtained by subtracting the image in the F555W filter, multiplied by a constant, from the image in the F657N filter; (b) image in the F814W filter; (c) image corresponding to the subtraction F814W - F657N, in magnitude scale; (d) image corresponding to F814W - F555W (equivalent to I-V), in magnitude scale, with the apparent dust disk near the center indicated by a gray dashed line.\label{HST}}
\end{center}
\end{figure*}

 \begin{figure*}
\begin{center}
   \includegraphics[scale=0.6]{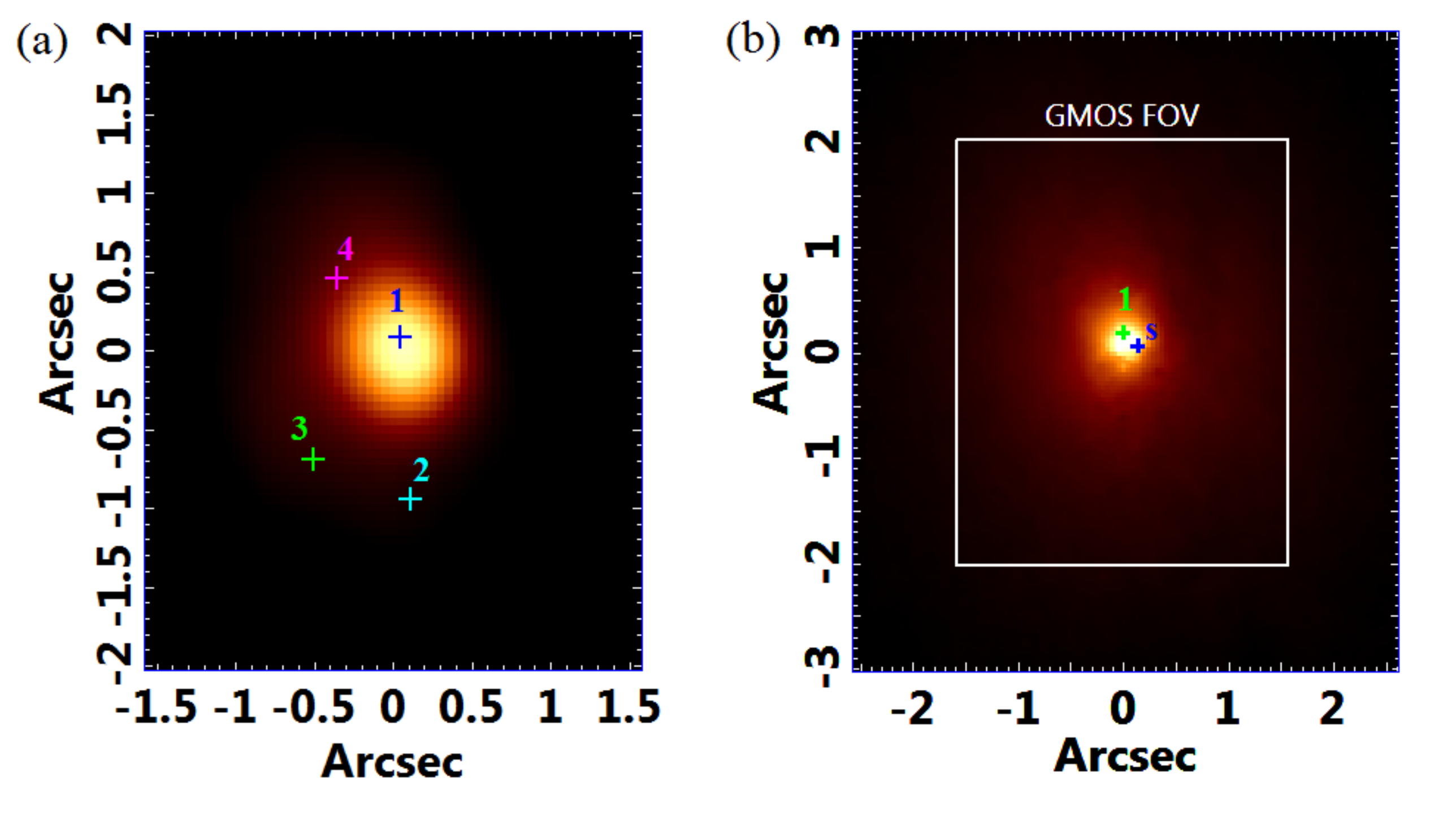}
  \caption{(a) Image of the stellar continuum of the data cube of NGC 6744 with the position of the four emitting regions indicated by crosses, and their sizes represent the uncertainty of 3$\sigma$.Note that Region 1 coincides with the stellar bulge center. (b) Image in the F814W filter (I band) with an indication of Region 1's position and the position of the blue source (S) detected in Fig.\ref{HST}d. \label{sourceregiao1}}
\end{center}
\end{figure*}

The spectral synthesis histogram of the nuclear region (Fig.~\ref{hist_star}a) shows that most of the flux fraction is due to 10$^{10}$ year old stellar populations and only $\sim 20$\% of the total flux fraction is due to populations of 10$^9$ years. However, the spectral synthesis of the circumnuclear region (Fig.~\ref{hist_star}b) reveals that the flux fraction is, approximately, double ($\sim 40$\%) for 10$^9$ year old stellar populations and the older ones are associated with a much lower flux fraction than in the nuclear region. There is no indication of younger populations or featureless continuum in these data. The fact that there is a greater flux fraction of the 10$^9$ year old population in the circumnuclear region suggests that this galaxy may have gone through a merger process that occurred one billion years ago, which generated this population (see Section \ref{sec8.4}).

The results provided by the spectral synthesis are subject to degeneracies. The Monte Carlo simulations described before can help to evaluate the uncertainties related to the S/N ratio but not the uncertainties due to such degeneracies. As a consequence, all the results presented in this section must be handled with caution. Therefore, although the exact ages determined for the stellar populations detected in the nuclear and circumnuclear spectra may not be totally reliable, the relative ages between these populations may be. One point in favor of this method is that the same technique was applied to analyze the nuclear and the circumnuclear spectrum, which makes the spectral synthesis results obtained for such spectra subject to the same kind of imprecisions. In addition, the observable differences between the stellar components of these two spectra make it clear that the stellar populations in these two regions are indeed different. Note that the absorption lines in the nuclear spectrum are deeper than the circumnuclear, which is evidence that there is a higher concentration of an old stellar population in the nuclear region compared to the circumnuclear region. Based on this, we conclude that, despite the degeneracies, a comparison between these results is a relevant procedure, worthy of being included in this work.

\section{Stellar Kinematics} \label{sec6}

In order to analyze the stellar kinematics of the data cube of NGC 6744, we applied the Penalized Pixel Fitting (pPXF) method \citep{cappellari}. It fits the stellar spectrum of an object with a combination of template stellar spectra from a determined base. In this case, the procedure was applied with the same base used to perform the spectral synthesis. The spectra of the base are convolved with a Gauss-Hermite expansion, providing the following parameters: stellar radial velocity ($V_*$), stellar velocity dispersion ($\sigma_*$), and the $h_3$ coefficient, which shows the profile deviations of the stellar absorption lines relative to Gaussian functions. Because this method was applied to all the spectra of the data cube, we obtained maps of those parameters (see Fig.~\ref{ppxf}). We found that a stellar continuum S/N ratio higher than 10 is necessary to obtain reliable kinematic parameters with pPXF. As the S/N ratio values of the analyzed data cube are $\sim 13$ on the edges of the FOV and $\sim 50$ close to the center of the stellar bulge, we concluded that the obtained maps of $V_*$, $\sigma_*$, and $h_3$ are reliable; though, the inaccuracies are higher at the edges of the FOV, due to the lower values of the S/N ratio (very close to the limit).

 Fig.~\ref{ppxf}(a) shows the $V_*$ map obtained with pPXF. There is a velocity bipolarity and the line of nodes has a PA = $3\degr \pm  2\degr$. The velocity curve, extracted along its line of nodes, seems to reveal a rotation in the inner region of the FOV. On the FOV edges, we see a lot of irregularities. The position of the kinematic center ($X_C=0\arcsec\!\!.00 \pm 0\arcsec\!\!.10$, $Y_C=0\arcsec\!\!.11 \pm 0\arcsec\!\!.10$) is compatible, at the 1$\sigma$ level, with that of Region 1, so we see an apparent stellar rotation around this region.

Fig.~\ref{ppxf}(b) shows the $\sigma_*$ map. We can see an evident decrease of $\sigma_*$ toward the kinematic center (Region 1), which can be called the $\sigma_*$-drop. The lowest detected value was $\sim$ 65 km s$^{-1}$, approximately at the position of Region 1.  

The $h_3$ map (Fig.~\ref{ppxf}c), though very noisy, reveals an apparent anticorrelation with the $V_*$ map. This indicates the presence of a probable stellar rotation superposed to a stellar background with $V_*$ close to zero.

\section{\textit{HST} images: the smoking gun}\label{hst}

The analysis of \textit{HST} images allows us a better and clearer visualization of structures in a larger FOV with higher spacial resolution. This certainly helps, in many cases, to identify those structures in the galaxy context and not only in the nuclear region.

The analyzed images were obtained with the Wide Field Camera 3 (WFC3) of the \textit{HST} on 2014 July 11 (F555W and F814W filters) and on 2015 July 11 (F657N filter). All images were retrieved from the \textit{HST} archive and were also rotated to the same orientation of GMOS data cubes in order to provide a more direct comparison. A rectangle representing the GMOS FOV was added (see Fig.~\ref{HST}), together with crosses indicating the position of the four main emitting regions. 

We subtracted the image in the F555W filter (multiplied by a constant) from the one in the F657N filter, in order to obtain an image representing the H$\alpha$ + [N \textsc{ii}] emission without the stellar continuum (Fig.~\ref{HST}a). Two emitting regions can be easily seen: the first one coincides with the position of Region 1 and the second, which is south from Region 1 and more diffuse, is close to the position of Region 2. The brightest central area extends slightly towards Region 4, and Region 3 cannot be detected.  

The F814W filter image (I band - Fig.~\ref{HST}b)  shows only a central bright area. This image was taken as a reference, as we observed in the GMOS data that the position of the center of Region 1 is compatible with the center of the stellar bulge (see Fig.~\ref{sourceregiao1}a). 

 Fig.~\ref{HST}(c) shows the image corresponding to the subtraction  F814W-F657N, in magnitude scale. The dark area represents the redder emission and the bright area indicates a more intense H$\alpha$ + [N \textsc{ii}] emission. We see again the two regions observed in Fig.~\ref{HST}(a). Besides that, there is a dark trail, in the western edges of those regions, indicating redder emission, possibly due to dust extinction. In this image, we can see more clearly the extension of Region 1 toward Region 4 and a little emission near Region 3.   
 
We obtained Fig.~\ref{HST}(d) by subtracting the image in the F555W filter from the one in the F814W, in magnitude scale. The resulting image shows a dark strip that crosses the FOV and bends toward Region 4, indicating that the spectra are redder there. Besides that, there is a compact blue source slightly west of Region 1 (for a better comparison of their positions, see Fig.~\ref{sourceregiao1}b). There is also, as is indicated in the image by the dashed gray line, a possible dark structure that crosses the compact source, which might be a dust disk. 
 
The projected distance between Regions 1 and 2, calculated from the \textit{HST} images, is $0\arcsec\!\!.91 \pm 0\arcsec\!\!.03$ (39.1 pc $\pm$ 2.0 pc), which is compatible with the one calculated from the GMOS data cube (see Table \ref{tabeladist}).

\section{Discussion} \label{sec8}

Despite being one of the brightest and closest Milky Way twins, NGC 6744 has been studied only briefly in the literature. In the present work, we saw that this object has a complex nuclear region from the standpoint of emission lines. The first notable characteristic is that, in the GMOS images, there are 4 relevant emitting regions. All the regions have emission lines ratios compatible with LINERs; though, Regions 2 and 3 apparently have higher ionization degrees than the others. Another important fact is that Region 3 appears more clearly in the [O \textsc{iii}]$\lambda$5007 line image, the morphology of this emission seems elongated toward Region 1 (which is also elongated), it is possible that this region is associated with an ionization cone. This hypothesis is completely acceptable within all involved uncertainties. If Region 3 has a higher ionization degree than Region 1 (if we consider the 1$\sigma$ level), this hypothesis is still valid if Region 3 has an electronic density lower than Region 1 or if it is being ionized by a region with a higher ionization degree, in this case, Region 2. Another hypothesis, ignoring the morphology in the [O \textsc{iii}]$\lambda$5007 line image, is that Region 3 is the ionization cone of Region 2 (which could alternatively be another AGN - see Section \ref{sec8.1}), and does not have any relation with Region 1.

On the other hand, Region 4, which seems to be in evidence in the [O \textsc{i}]$\lambda$6300 image, shows a lower ionization spectrum, so it can be part of the NLR of the AGN in Region 1.

In the \textit{HST} images analyzed here, we see that only the areas associated with Regions 1 and 2 are easily identified in the image of the extended emission of H$\alpha$ + [N \textsc{ii}] (Fig.~\ref{HST}a). The projected distance between those regions determined from the \textit{HST} images is compatible with the one determined from the GMOS data cube ($\sim$ 0\arcsec\!\!.9).

There is an emission that extends from Region 1 toward Region 4, seen in Fig.~\ref{HST}(c), evidencing the relation between those regions and indicating that Region 4, as said before, seems to be part of the NLR of the AGN. 

The I-V image from the \textit{HST} (see Fig.~\ref{HST}d) clearly shows a point-like source, with blue emission. This source does not seem to be associated with stars, since the spectral synthesis results detected no young stellar population. As the position of this blue source is compatible, within the uncertainties, with Region 1, we conclude that the most probable scenario is that this source is the AGN. Although we did not detect featureless continuum emission in the spectral synthesis results, this blue source can be due to a featureless continuum, as this AGN has low luminosity and this emission is diluted in GMOS PSF (making it harder for the spectral synthesis to detect the featureless continuum).

We see a red trail west of Region 1 and a little above Region 4. This reddening can be explained by the presence of dust \citep{fisher}. We can also see a compact structure of dust crossing the blue compact source. This may indicate a gas and dust disk, which might be associated with the process of feeding of the AGN (see Fig.~\ref{HST}d). We note that Region 3, identified as an ionization cone, is approximately perpendicular to this dust and gas disk. Besides that, it is not possible to identify an opposite emission of Region 3 (which might be the other side of the ionization cone), as this area is strongly obscured by dust.

\subsection{Photoionization or Shocks?}\label{sec8.1}

When one has an evident LINER emission in the studied object, it is natural to question whether this emission is due to photoionization by a central AGN or it is the result of shock waves caused, for example, by supernovae. In the present case, we have some different regions with different ionization degrees that deserve to be properly investigated, considering the properties of the line emission and its morphologies. 

By analyzing the electronic densities, determined from the [S \textsc{ii}]$\lambda\lambda$6716, 6731 doublet, we see that, in all regions, the obtained values are considerably low, with Region 2 being the densest region with electronic density between 230 and 409 cm$^{-3}$ (Table \ref{tableetc}). Besides that, Regions 2, 3, and 4 have very low temperature limits, lower than $10^4$ K - certainly compatible with the idea of being clouds photoionized by a central AGN (located in Region 1).

The emission lines of all observed regions are quite narrow (typically with FWHM of $\sim$ 150 km s$^{-1}$). This leads naturally to the following question: are those gas velocities sufficiently high for a model of heating by shock waves to explain the detected emission spectrum? We then made some tests using the Mappings III \citep{allen08} library, with only solar abundance data and with the velocities observed here. We saw that those velocities are not sufficient to reproduce the emission line ratios in a shock wave scenario. An important fact to be considered is that the spectral synthesis did not detect stellar populations younger than one billion years, invalidating the hypothesis of the existence of regions with concentrations of supernovae remnants. With this result, we can discard this hypothesis, with the only remaining models involving AGN photoionization or post-AGB stars (\citealt{binette} and \citealt{cid11}). Since NGC 6744 has a central compact blue source, it is unlikely that the ionization mechanism is due to post-AGB stars.  

\subsection{One or Two AGNs?}

As mentioned before, four regions with relevant emission were observed in this data cube. There are many possible scenarios, involving even two AGNs, that can explain the coexistence of those emitting regions and that depend on the uncertainties considered in the calculation of the emission line ratios. 

If we consider the 3$\sigma$ level, the emissions of all regions are compatible. Under these circumstances, a possible scenario assumes that there is only one source ionizing all those regions, Region 1, since it has higher luminosity and is an AGN (because of the blue compact source observed in the \textit{HST} image). Then, Region 1 would be the central AGN with LINER emission and the other regions would be part of its NLR. 

We also hypothesize that Region 2 is another AGN that is not clearly detected in the \textit{HST} images. This region can have an ionization degree higher than that of Region 1, supporting the hypothesis of a merger event with a galaxy without a defined stellar nucleus around its central black hole. In this case, the central black holes of both galaxies did not coalesce yet. 

\subsection{A Very Low-luminosity AGN, Fossil of a More Luminous Phase}

In order to explain why an AGN of such low luminosity can excite clouds that are so distant with an ionization parameter similar to the central region, we consider the hypothesis called "AGN echo," proposed, for example, for SDSS J094103.80+344334.2 (\citealt{lintott} and \citealt{keel}). In this case, Regions 2 and 3 would be ionized by an emission coming from  Region 1, which would have had a higher luminosity and ionization degree in the past. However, over time, the luminosity of this emission decreased (with lower feeding rate, for example) and reached the current ionization degree (lower than before). This hypothesis is consistent with the width of the lines that we observed, since an AGN with lower activity usually shows fewer outflows, which broadens the emission lines. In this scenario, the other regions, which are part of the NLR of the AGN, are still receiving (except Region 4, the closest one) the old emission from this AGN, when it had a higher luminosity and ionization degree.

The H$\alpha$ luminosity of the nucleus is 3.4 $\times$ 10$^{37}$ erg s$^{-1}$, about 10 times higher than in the other regions (Table \ref{tableetc}). Considering that the representative $\sigma_*$ value of the bulge of this galaxy is $\sim$ 95 km s$^{-1}$ (excluding the $\sigma_*$-drop observed in the pPXF maps - see Fig.~\ref{ppxf}), we conclude that the black hole has a mass of $\sim$ 6 $\times$ 10$^6$ M$_{\bigodot}$ \citep{gultekin}. Assuming the bolometric luminosity given by \citet{netzer} to narrow line objects (type 2), the case B recombination, where H$\alpha$/H$\beta$= 2.87, a temperature of T= 10$^4$ K in the low density limit \citep{osterbrock} and the H$\alpha$ luminosity from Region 1, we obtain a bolometric luminosity of L$_{bol}$ $\sim$ 3.2 $\times$ 10$^{40}$ erg s$^{-1}$ and an Eddington ratio of $\sim$ 4 $\times$ 10$^{-5}$. This is an Eddington ratio typical of LINERs \citep{ho08,ricci}.

\subsection{Evidence of a Merger}\label{sec8.4}

There is evidence of a merger, revealed by the spectral synthesis results, which showed only two significant star formation episodes: the first one occurred 10$^{10}$ years ago, during the formation of the galaxy, and the second one occurred 10$^9$ years ago.

The spectral synthesis was applied to two spectra: one of the nucleus and one of the circumnuclear region. There is a significant difference between these regions: the higher concentration of the 10$^9$ year old stellar populations in the circumnuclear region and the higher concentration of the older stellar populations (10$^{10}$ year old) in the nucleus. This difference can be noticed in the spectra  because the nuclear spectrum has deeper absorption lines. The  difference of the fluxes associated with $10^9$ year old stellar populations between the nuclear and circumnuclear regions suggests that this galaxy went through a merger one billion years ago, which resulted in a circumnuclear starburst and the formation of this stellar population. After this possible merger, there was no other relevant star formation episode in the central region of the galaxy. In fact, as \citet{fisher} verified, NGC 6744's bulge is similar to an elliptical galaxy, with little star formation and old stellar populations. It is important to notice, however, that NGC 6744 does not show evidence of a double stellar nucleus. As mentioned before, despite the degeneracies of the spectral synthesis results, the observable differences between the stellar components of the nuclear and circumnuclear spectra, together with the fact that the same procedure was applied to these spectra (and, therefore, the results are subject to the same types of imprecisions), make this analysis reliable.

Since NGC 6744 is classified as SAB(r)bc, with a prominent disk, it is not an expected outcome of a major merger. However, studies have shown that, under certain circumstances, major mergers can result in late-type galaxies (\citealt{barnes2002,Springel} and \citealt{robertsonbrant}). Usually, this requires that the galaxies involved in the merger have a high gas fraction. Therefore, although scenarios involving major mergers do not appear to be the most likely to explain the results obtained for the data cube of NGC 6744, they cannot be discarded. On the other hand, minor merger scenarios certainly deserve special attention. Studies about such scenarios indicate that, unlike major mergers, minor mergers, with high mass ratios, result only in modest star formation in the central region of the galaxy (\citealt{taniguchi} and \citealt{cox}). Although the flux fraction associated with the stellar population probably related to the merger (with 10$^9$ years and intermediate metallicity) represents $\sim 40\%$ of the total stellar flux in the circumnuclear region of the data cube of NGC 6744, we cannot precisely evaluate how extended this starburst event was due to our limited FOV.

\subsection{Gas Kinematics}

From the H$\alpha$ velocity map (Fig.~\ref{vel_ha}), one can notice that Regions 2 and 3 are in blueshift relative to Region 1 and have very close and compatible (at $1\sigma$ level) radial velocity values, suggesting that those regions are part of the same structure, which is different from the one corresponding to Regions 1 and 4. The H$\alpha$ velocity map also shows that there is a gas rotation around Region 1, but there are many irregularities on the edges of the FOV, which can be due to the presence of other kinematic phenomena, such as inflows or outflows, for example. It is also possible that those deviations from a simple Keplerian rotation are consequences of the probable merger that has occurred in this galaxy, as discussed above, or even the result of imprecisions of the fits of the Gaussian functions to the H$\alpha$ emission lines in the areas with lower A/N ratios (which are, indeed, the edges of the FOV). The channel maps in Fig.~\ref{channelmaps} show the same rotation pattern as that observed in the H$\alpha$ velocity map. However, we see that the PA of the line of nodes in the region with blueshifted gas seems to be different from the PA of the areas with redshifted gas. Again, this can be a consequence of either the presence of different kinematic phenomena or the fact that the system is still going through the results of the merger.

\subsection{Stellar Kinematics}

Stellar kinematic analysis revealed a rotation around Region 1 with a PA considerably different from that of the gas rotation. Assuming that there is a gas disk and a stellar disk, we conclude that structures are not coplanar. As seen in the H$\alpha$ velocity map, there are irregularities at the edges of the FOV of the $V_*$ map that could be due to either consequences of a merger or imprecisions of the pPXF method, because those areas have low S/N ratios (close to the lower limit of 10).

We detected a $\sigma_*$-drop toward the center of the bulge (Region 1). The decrease of $\sigma_*$ toward the center is explained in the literature as the result of the presence of young stellar populations that herd the kinematics of their progenitor clouds that had low velocity dispersions \citep{woz03,all05,all06}. However, in this case, the spectral synthesis detected very low flux fractions associated with young stellar populations; so this hypothesis is not valid. A more exotic scenario to explain this $\sigma$-drop pattern may be that the possible merger involved another galaxy without a well defined stellar nucleus (an Sm galaxy, for example). Thus, this event would result in a starburst of which the effect would be more significant in the circumnuclear region  but would not affect the stellar kinematics of the central region of the original nucleus of NGC 6744, as no coalescence would occur with another galactic nucleus. In this scenario, the observed $\sigma_*$-drop would simply be related to the original nucleus of NGC 6744, while the $10^9$ year old stellar population would be a result of this merger. 

The $h_3$ coefficient map showed that there is an apparent anticorrelation with the stellar velocity map, which suggests that the stars that are in rotation  are superposed to a background of stars with velocity close to zero (the bulge stars).

\section{Conclusions}\label{sec9}

In this work, we analyzed a data cube of the nuclear region of NGC 6744 obtained with GMOS/IFU, together with \textit{HST}/WFC3 images. The main conclusions of this work are the following:

(1) There are four relevant emitting regions in the center of NGC 6744. Region 1 is coincident (within the uncertainties) with the center of the stellar bulge and, also, with the kinematic center of the gas and stellar rotations. All regions have emission line ratios compatible with LINERs, but Regions 2 and 3 seem to have higher ionization degrees when compared with the other two.

(2) An I-V image obtained with \textit{HST} revealed the emission of a compact blue source; as this galaxy does not show any evidence of young stellar populations in its central region, our interpretation is that this fact confirms the existence of an AGN of low luminosity (Region 1).

(3) The most probable scenario that explains the coexistence of the four observed regions assumes that Regions 2, 3, and 4 are part of the NLR of the AGN (Region 1). Region 3 seems to be an ionization cone of this AGN, due to its considerable emission in the [O \textsc{iii}]$\lambda$5007 line and to its morphology. The fact that Regions 2 and 3 have higher ionization degrees than Region 1 can be justified by the hypothesis of them being ionized by an emission coming from the AGN in Region 1, when it was in a more luminous phase.

(4) We detected a compact structure of dust crossing the galactic nucleus in the I-V image from the \textit{HST}. This seems to indicate a probable gas and dust disk that may be feeding the AGN. Region 3, which we identified as a possible ionization cone, appears to be perpendicular to this disk of dust and gas.

(5) The spectral synthesis was applied to two spectra of the GMOS data cube, one from the nucleus (Region 1) and the other from the circumnuclear region. This revealed that this galaxy has passed for two main star formation periods: one during the formation of the galaxy and another, 10$^9$ years ago. After this period, there were not any significant stellar formation events. This last starburst can be associated with a significant merger that happened one billion years ago.

(6) There is a gas rotation around Region 1, as seen in the H$\alpha$ velocity map. Channel maps of the main emission lines of the GMOS data cube revealed that the PA of the line of nodes of the blueshifted gas is different from that of the redshifted gas, which may be the result of inflows or outflows of gas or the result of the possible merger.

(7) The stellar kinematic analysis with the pPXF method revealed that there is a stellar rotation around Region 1, but the PA is different from the one determined for the gas rotation. This stellar rotation seems to be superposed to a background composed of stars with velocity values close to zero, which can be seen by the anticorrelation between the $h_3$ coefficient map and the V$_*$ map. The $\sigma_*$ map showed a $\sigma_*$-drop toward the center. A relatively exotic scenario to explain this involves a merger process with a galaxy without a well defined stellar nucleus. 

\acknowledgments
This work is based on observations obtained at the Gemini Observatory (processed using the Gemini IRAF package), which is operated by the Association of Universities for Research in Astronomy, Inc., under a cooperative agreement with the NSF on behalf of the Gemini partnership: the National Science Foundation (United States), the National Research Council (Canada), CONICYT (Chile), the Australian Research Council (Australia), Minist\'erio da Ci\^encia, Tecnologia e Inova\c{c}\~ao (Brazil) and Ministerio de Ciencia, Tecnolog\'ia e Innovaci\'on Productiva (Argentina). This work is also based on observations made with the NASA/ESA Hubble Space Telescope, obtained from the Data Archive at the Space Telescope Science Institute, which is operated by the Association of Universities for Research in Astronomy, Inc., under NASA contract NAS 5-26555. These observations are associated with the following programs: 13364 and 13773. We thank CNPq (Conselho Nacional de Desenvolvimento Cient\'ifico e Tecnol\'ogico), CAPES (Coordena\c{c}\~ao de Aperfei\c{c}oamento de Pessoal de N\'ivel Superior) and FAPESP (Funda\c{c}\~ao de Amparo \`a Pesquisa do Estado de S\~ao Paulo), under grant 2011/51680-6, for supporting this work. This research also has made use of the NASA/IPAC Extragalactic Database (NED), which is operated by the Jet Propulsion Laboratory, California Institute of Technology, under contract with the National Aeronautics and Space Administration.

\bibliographystyle{yahapj}
\bibliography{references.bib}


\end{document}